# Fresnel-type Solid Immersion Lens for efficient light collection from quantum defects in diamond

SUNGJOON PARK,[1] GYEONGHUN KIM,[2] KIHO KIM,[1] AND DOHUN KIM[1,*]

[1]*Department of Physics and Astronomy, and Institute of Applied Physics, Seoul National University, Seoul 08826, Korea*
[2]*Department of Physics, Duke University, Durham, NC 27708, USA*
*\*dohunkim@snu.ac.kr*

**Abstract:** Quantum defects in diamonds have been studied as a promising resource for quantum science. The subtractive fabrication process for improving photon collection efficiency often require excessive milling time that can adversely affect the fabrication accuracy. We designed and fabricated a Fresnel-type solid immersion lens using the focused ion beam. For a 5.8 µm-deep Nitrogen-vacancy ($NV^-$) center, the milling time was highly reduced (1/3 compared to a hemispherical structure), while retaining high photon collection efficiency (> 2.24 compared to a flat surface). In numerical simulation, this benefit of the proposed structure is expected for a wide range of milling depths.

## 1. INTRODUCTION

Photoactive quantum defects have garnered considerable attention as key components in quantum technologies such as quantum computing[1–6], quantum sensing[7–15], and quantum communication[16–19]. Among them, the negatively charged nitrogen-vacancy ($NV^-$) center has been extensively studied owing to its long coherence time[20–22], possibility of achieving a multi-qubit register with nearby nuclear spins[23–26], and capability of position-controlled generation[27–32]. The quantum states of electrons in the $NV^-$ center can be distinguished by optical measurement[33] where the cumulative collection of light emitted from the $NV^-$ center in a range of wavelengths from 600 nm to 800 nm shows spin state-dependent photoluminescence[34].

One of the obstacles in detecting the spin state of the $NV^-$ center with high fidelity is the low photon collection efficiency originating from the high refractive index of diamond of approximately 2.4. Various methods have been developed to improve the photon collection efficiency, such as pillar structures[35–37], photonic crystals[38–40], solid immersion lenses (SIL)[41–43], and meta lenses[44]. Subtractive fabrication-based solid immersion lenses have been widely used because of their addressability of deep emitters, easy targeting, and stability in chemical processes. SIL is also widely used for low-temperature experiments, where additional photon loss across the cryogenic chamber demands tools for photon collection enhancement, as well as the temperature stability of the structure. However, the milling time required for the conventional hemispherical solid immersion lens (h-SIL) increases with the cube of the emitter depth. Moreover, the increased milling time can increase the uncertainty of fabrication owing to a charging effect and a stage drift over time, as well as an unintended strain effect on $NV^-$ centers[45]. Thus, a new approach that reduces the milling time and maintains high photon collection efficiency is required.

In this study, we designed and fabricated a Fresnel-type SIL using a focused ion beam (FIB) technique for a single $NV^-$ center in diamonds. The structure is composed of concentric spherical shapes connected by cones to not only enhance the photon collection efficiency compared with a flat surface but also reduce the milling time cost. Using this method, we showed that the photon collection efficiency is enhanced to 2.24 times higher than that of a flat surface under an oil immersion lens, and the milling time is reduced to less than 1/3 of that of a hemispherical solid immersion lens (h-SIL). Our design can be used for high-fidelity optical readout of spin states[46] related to quantum computing and optically addressed quantum networks, especially for photo-luminant quantum defects far from the surface of the host material.

## 2. EXPERIMENTAL IMPLEMENTATION

We used a 2 × 2 × 0.5 mm type-IIa single-crystal diamond grown by the chemical vapor deposition method containing nitrogen impurities below 5 ppb (element six). To increase the number of NV$^-$ centers, the sample was irradiated with electrons at an acceleration energy of 2 MeV for a few minutes and then annealed at 1200 °C for 2 hours under high vacuum conditions (≈3×10$^{-7}$ Torr). After annealing, the sample was cleaned using triacid (a 1:1:1 mixture of sulfuric, nitric, and perchloric acid) [47] over 1 h. For the FIB fabrication, we used a 30 keV-gallium FIB system (FEI Helios G5), where three-dimensional milling was processed with stream files containing Cartesian coordinates of beam positions and beam dwelling times [43]. The milling rate was calibrated based on the measured ion-beam current. To reduce the charging effect of Ga+ ions on the surface of the diamond, we coated the sample with 25 nm Cr using an e-beam evaporator (SORONA SRN-200). After the FIB milling process, the Cr coating was removed with a Cr etchant (TRANSENE Cr ETCH 905N), and the sample was cleaned with triacid.

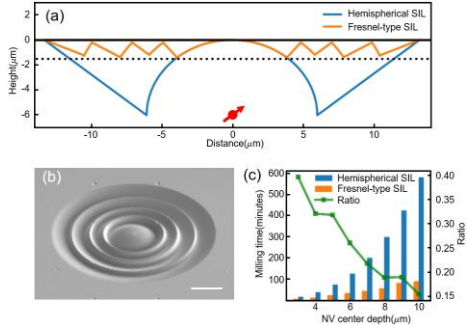

**Fig. 1. (a)** The cross-section of a conventional hemispherical SIL (h-SIL, blue) and the Fresnel-type SIL (orange) for addressing an NV$^-$ center 6 µm below the surface. The dashed black line is a preset milling depth limit $d_0$. **(b)** Scanning electron microscope image of a representative Fresnel-type SIL. The scale bar is 5 µm. **(c)** Comparison between required milling times to fabricate h- (blue) and Fresnel-type (orange) SIL as a function of depth of the NV$^-$ center $d$.

Figure 1a shows the profiles of the h-SIL (blue) and Fresnel-type SIL (orange) designed for an NV$^-$ center 6 µm below a diamond surface. In both profiles, the rays from the NV$^-$ center are perpendicular to the arc. The major difference between the two designs is that the volume that needs to be removed increases by the cube (square) of the depth of the NV$^-$ center for h-SIL (Fresnel-type SIL). The design of the current Fresnel-type SIL also differs from that of the conventional Fresnel lens in that the former aims to minimize the disturbance in the radiation path from the light source, while the latter guides collimated incident light to a focal point.

The detailed design strategy for our Fresnel-type SIL is described in S1 in the Supporting Information. Briefly, the radius of the nth sphere was recursively determined for a given maximum milling depth, $d_0$. First, for constructive interference between photons emitted from the NV$^-$ center, $R_n$ satisfies the recursion relation

$$R_n = R_{n-1} + \frac{i \times \lambda}{|n_0 - n|} , \qquad (1)$$

where $i$ is a positive integer, $\lambda$ is the wavelength emitted from the NV⁻ center, $n_0$ ($n$) is the refractive index of immersion oil (diamond). Next, $i$ was determined to have a maximum value that satisfies

$$R_n \cos\theta = d \quad \text{or} \quad d - l, \qquad (2)$$

$$R_{n-1} \cos\theta > d - d_0 , \qquad (3)$$

where $d$ is the depth of the NV⁻ center, and $l$ is the technical depth limit of the FIB machine caused by proximal milling. Using Eqs. (1)–(3), we limit the milling depth and decrease the adverse effects of photon collection related to destructive interference. Figure 1b shows a scanning electron microscope image of a representative Fresnel-type SIL structure, the performance of which is discussed later.

Figure 1c compares the required milling times for Fresnel-type and h-SIL targeting NV⁻ centers at depths ranging from 3 to 10 µm at an identical milling rate and condition (for Fresnel-type SIL, we used $\lambda = 700$ nm, $d_0 = 1.5$ µm, $l = 0$), calculated from the generated stream files. Starting from the reduced milling time of 60% for the NV- center at a depth of 3 µm compared with h-SIL, the Fresnel-type SIL design enables affordable FIB milling time even for deeper NV⁻ centers. The demonstrated reduction in milling time of the Fresnel-type SIL lowers the time cost and the risk of resulting in an unintentional off-centered structure owing to stage drift and other mechanical instability over time.

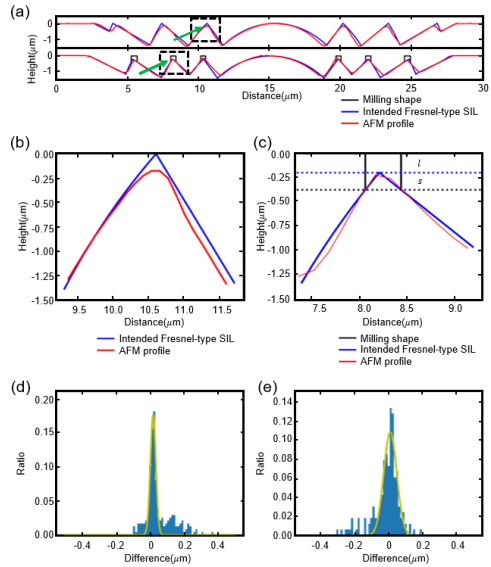

**Fig. 2. (a)** Topography obtained from atomic force microscopy (red) of fabricated Fresnel-type SILs overlaid with intended lens profile (blue) and milling profile (black). Top panel: the first fabrication method where the milling shape naively follows the design. Bottom panel: method to compensate for the proximal milling effect where the region near the apex of the structure is intentionally omitted from the milling. The magnified profiles near the apexes (dashed black boxes in (a)) are shown in **(b)** for the first and **(c)** second approaches, respectively. **(d)** and **(e)**

show the histograms of the vertical milling errors obtained profile from **(b)** and **(c)**, respectively, with Gaussian fit.

The sharp edge structures of the Fresnel-type SIL were realized through the two milling strategies depicted in Fig. 2a around NV⁻ centers whose positions were pre-calibrated by confocal microscopy [43] (see Fig. S2 of Supporting Information). After FIB milling process, the topography was examined using atomic force microscopy (AFM). In the first strategy, the milling profile followed the intended Fresnel-type SIL shape (Fig. 2a, top panel). The resultant apex of the edges shows a deviation of up to several 100 nm (Fig. 2b) from the design owing to the proximity effects. As described in S3 of the Supporting Information, we observed a moderate photon collection efficiency enhancement with this approach (~1.5) compared to a flat surface. The other strategy is to place a rectangular profile near the apex of the arc (bottom panel of Fig. 2a), where milling is blanked[48,49]. For example, for $l$ = 200 nm (blue dotted line in Fig. 2c), the spaces near the edges and depths down to 380 nm (black dotted line in Fig. 2c) are omitted from the milling profile. Using this strategy, we fabricated the Fresnel-type SIL around a 5.8 µm deep NV⁻ center, as shown in Fig. 1b. As a result, the AFM profile (Fig. 2c) more closely resembles the intended lens design. Figure 2d (2e) shows a histogram of the differences between the AFM profile and Fresnel-type SIL design for the first (second) milling strategy. From the Gaussian fit, we obtained standard deviations < 40 nm, which are much smaller than the wavelengths considered for this experiment (> 532 nm). In the first milling strategy (Fig. 2d), the side peak near 120 nm likely stemmed from proximal milling. This flaw may contribute to additional refraction and destructive interference, reducing the photon collection efficiency compared with an ideal case.

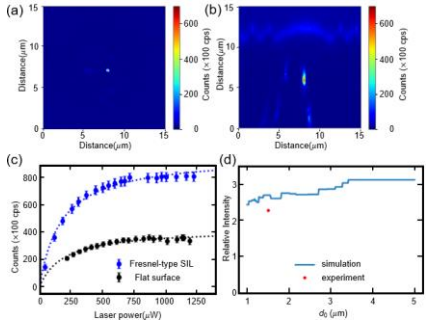

**Fig. 3.** **(a)** Lateral and **(b)** vertical cross-sectional confocal microscopy image of the single NV⁻ center with the fabricated Fresnel-type SIL **(c)** Photon counts as a function of laser power for the NV⁻ center with (blue) and without (black) Fresnel-type SIL. The dashed curves are least-square fits to the saturation model for a single-photon source. **(d)** The numerical simulation of the expected photon collection efficiency enhancement for the Fresnel-type SIL as a function of $d_0$. The red point marks the experimental result.

### 3. PHOTON COLLECTION EFFICIENCY

We analyzed the performance of the Fresnel-type SIL fabricated using the second milling strategy described above. A confocal scan of a 15 × 15 µm² area around a 5.83 µm deep NV-center was measured in the vertical (lateral) cross-sectional plane, as shown in Fig. 3a (3b), using an excitation laser with a wavelength of 532 nm and a power of 475 µW. The light emitted by the NV⁻ center was collected using an oil immersion objective with a numerical aperture of 1.3. To quantify the enhancement in photon collection efficiency, we measured the photoluminescence from NV⁻ centers under the Fresnel-type SIL and a flat surface with varying laser power (Fig. 3c). In both cases, the single quantum emitter characteristics were confirmed by Hanbury Brown and Twiss (HBT) measurement [50] (see Fig. S2f of Supporting Information). For a Fresnel-type SIL (flat surface), the photon counts per second (cps) reach >

80 kcps (< 40 kcps), which shows a significant enhancement in photon collection efficiency. The dashed curves in Fig. 3c fit the saturation model $F = F_{sat}I/(I_{sat}+I)$ [51], where Fsat is the saturated photoluminescence, and $I_{sat}$ is the saturation laser power. We obtain $F_{sat}$ = 95.5 (42.0) kcps, $I_{sat}$ = 175 (190) µW with the Fresnel-type SIL (flat surface) showing that the enhancement in photon collection efficiency for the Fresnel-type SIL is > 2.24. We also expect that the improvement will be greater using an air objective because the critical angle for total reflection is smaller.

Figure 3d shows the finite difference method (FDM) simulation (see S5 in the Supporting Information for more details). When $d_0$ is limited to extremely small values, complex patterns near the surface of the diamond with sizes comparable to the wavelength of the emitted light lead to a significant reduction in the photon collection efficiency. However, for nominal $d_0$ of the order of 1 µm, a significant photon collection efficiency of > 2.5 can be obtained. The numerical simulation also shows that the enhancement in the photon collection efficiency of our Fresnel-type lens approaches that of the h-SIL as $d_0$ > 3.5 µm. Currently, although the experimental result is close to the theoretical expectation, the remaining difference between the experiment and theory may stem from the imperfect realization of the edge structure and reflection at the diamond surface [52].

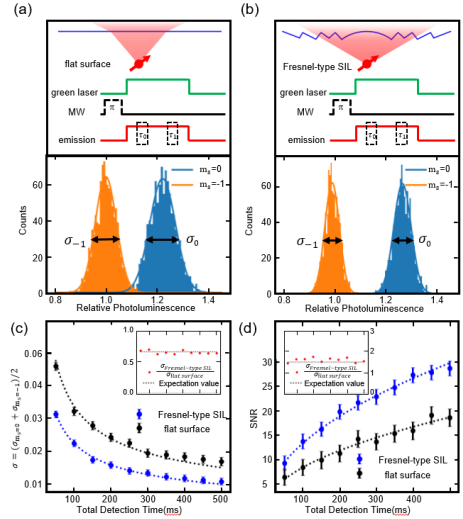

**Fig. 4.** Histograms of photoluminescence from the NV[-] center without **(a)** and with **(b)** the Fresnel-type SIL. The top panels show schematic cross section of each design and pulse sequence used for the experiment. Variation of **(c)** average standard deviation ($\sigma$) and **(d)** signal to noise ratio (SNR) as a function of total detection time. The insets show nearly constant $\sigma$ and SNR as a function of total detection time.

## 4. SPIN READOUT

Using the fabricated Fresnel-type SIL, we showed an improvement in the signal-to-noise ratio (SNR) for determining the spin states (spin quantum number $m_s$ = 0 vs. $m_s$ = -1) of the NV-center. To prepare the excited state with $m_s$ = -1, we manually positioned a gold wire near the milled structure and applied a π-pulse calibrated from a microwave-driven Rabi oscillation. Figures 4a and 4b show histograms of the normalized spin-state-dependent photoluminescence from the NV[-] center with a flat surface and the Fresnel-type SIL, respectively. The histograms were based on 1000 records, each of which consisted of accumulated photon counts in $10^6$ repeated readout stages with a time window of 50 ns (total detection time = 50 ns × $10^6$ = 50 ms). See Fig. S4 in the Supporting Information for more details on the choice of experimental conditions, such as the timing of readout windows in gated photon counting.

From the Gaussian fit to the data, we obtained the uncertainty in state measurement $\sigma_{0\,(-1)}$ for the spin state with $m_s=0\,(-1)$. Figure 4c shows the average uncertainty $\sigma = (\sigma_0 + \sigma_{-1})/2$ as a function of the total detection times for both the flat surface and the Fresnel-type SIL. Limited by photon shot noise, $\sigma$ is inversely proportional to the square root of the total detection time (dashed curves in Fig. 4c). The enhancement in the standard deviation was constant over a wide range of detection times, as shown in the inset of Fig. 4c. Figure 4d shows the signal-to-noise ratio ($\langle SNR \rangle = (N_0 - N_1)/\sqrt{N_0 + N_1}$ [53]).

## 5. CONCLUSION

In conclusion, we demonstrated a Fresnel-type SIL that enables efficient light collection from photoluminescent quantum defects in high-refractive-index materials and facile fabrication within an affordable milling time. The FDM simulation showed that the photon collection efficiency was > 2.5, and the milling time was expected to be reduced to 15% of that required for h-SIL for $NV^-$ centers 10 µm below the diamond surface. With the Fresnel-type SIL, we achieved the photon collection efficiency enhancement > 2.24 compared to that of the flat surface and needed only 30 min for a net milling time for $NV^-$ center 5.83 µm below the diamond surface (for h-SIL, it may take more than 100 min). Our experiment broadens the possibility of using shallow solid structures to address deep quantum defects, which can be used in various quantum information experiments.

**Acknowledgments**
We thank to Byungnam Kim and Youn-Mook Lim at the Korea Atomic Energy Research Institute (KAERI) for the technical support in electron irradiation. We thank to Youngwoo Jeong at the Korea Institute of Science and Technology Advanced analysis and data center (KIST AAC) for the technical support in ion beam milling. We thank to M.H. Abobeih and T.H. Taminiau at QuTech, Delft University of Technology for the advice about conducting layer coating.

**Disclosures**
The authors declare no conflicts of interest.

# Fresnel-type Solid Immersion Lens for efficient light collection from quantum defects in diamond

SUNGJOON PARK,[1] GYEONGHUN KIM,[2] KIHO KIM,[1] AND DOHUN KIM[1,*]

[1]Department of Physics and Astronomy, and Institute of Applied Physics, Seoul National University, Seoul 08826, Korea
[2]Department of Physics, Duke University, Durham, NC 27708, USA
*dohunkim@snu.ac.kr

**Supporting Information**

### S1. Inductive designing process of the Fresnel-type SIL

From Eqns. (1)–(3) in the main text, we have created candidate designs for Fresnel-type SIL. The first arc radius was chosen as the depth of the NV$^-$ center from the surface. Moreover, the radius of the second arc is generated using Eq. (1), and the radius of the first arc. Starting with $i$=1 (Fig. S1. a red dotted line(i)), we increased $i$ until Eq.(3) was violated (Fig. S1. Blue dotted line(iii)). Then, we select one, which is right before violation, as the radius of the second arc radius (Fig. S1. a yellow line(ii)). The third arc radius was generated using Eq. (1) and radius of the second arc, as previously explained.

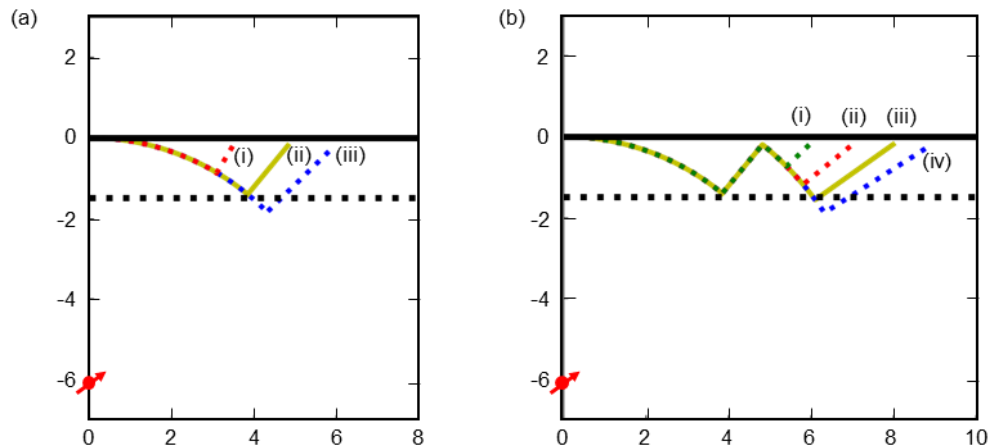

**Figure S1. (a)** and **(b)** show candidate designs. The profile candidates of the first cone are shown in **(a)**, and those of the second are shown in **(b)** as (i)–(iv). In both cases, the yellow line graph represents the final design determined by the constraints($d_0$).

### S2. Milling process for the Fresnel-type SIL.

To address a single NV$^-$ center in bulk diamond, we used a home-made confocal microscope (CFM) before FIB milling (Fig. S2. a-c). Before taking it, we milled corner dots serving as align marker later. After removing the conducting layer prior to FIB milling, we observed a dark spot at the dots in the CFM (Fig. S2. b red circles). We also found an in-depth single NV- center in the CFM (Fig. S2. c red circle). After milling, we obtained these CFM again, which shows a clear change in the photon collection efficiency (Figure S2. d and e). To quantify this and confirm single photon emitter characteristics, we measured the auto-correlation (HBT) measurement (Figure S2. f). The value of auto-correlation at zero-delay time ($g^2(0) \approx 0.017$) indicates a very low background signal[1].

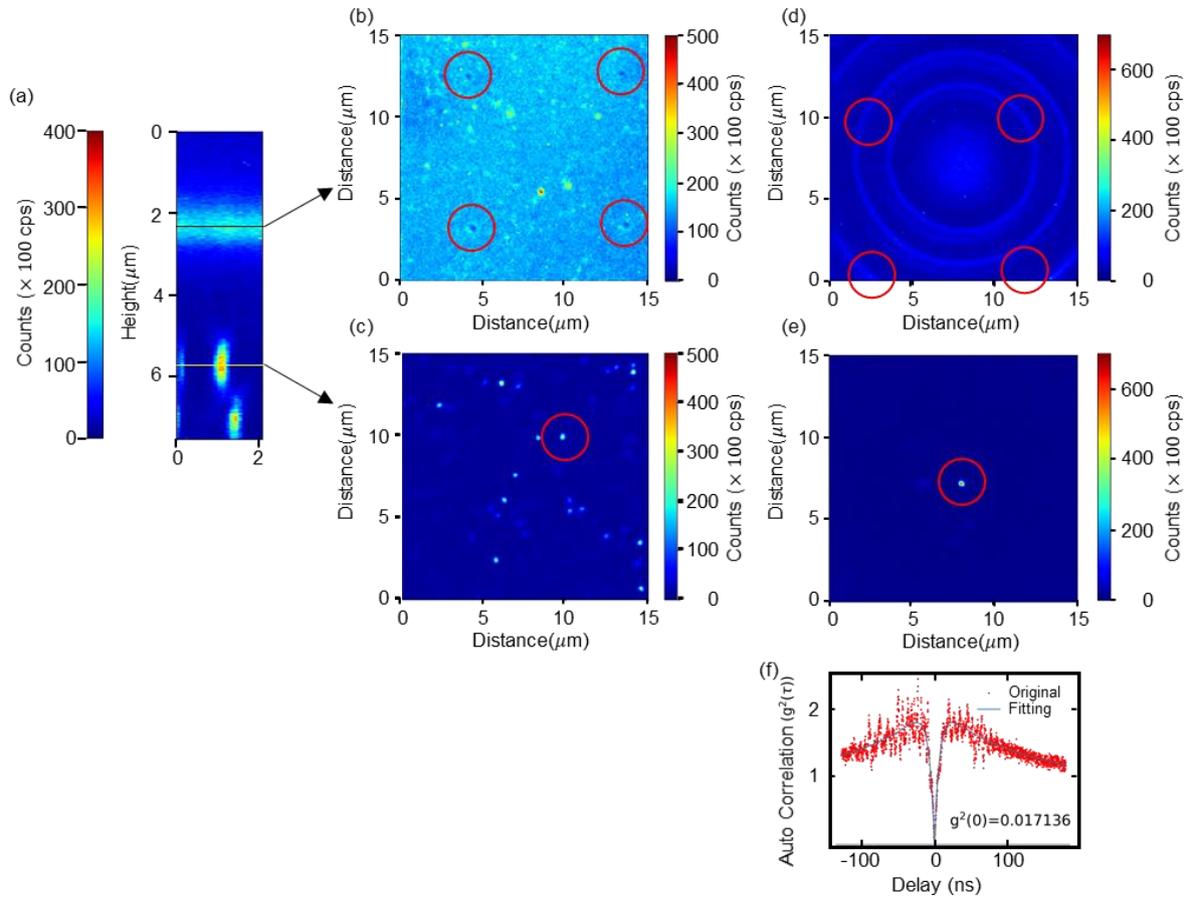

**Figure S2. (a)** and **(b)**, **(c)** show confocal microscopes (CFM) in the cross-sectional and lateral sectional planes, respectively, before milling the Fresnel-type SIL. **(b)** is taken near the surface where the FIB grid is made, and **(c)** is taken at the target NV- center's depth. After FIB milling, the CFM was obtained from the same region(**(d)** and **(e)**). The red circles show the positions of the grid dot and NV center. **(f)** HBT measurements taken from the same NV- center.

### S3. Different milling strategy for the Fresnel-type SIL

We tested two strategies for strategy (Fig. 2). The second strategy is described in this study. The CFMs resulting from the first strategy are shown in Fig. S3. Even though it has a blunt end (Fig. 2b), the photon count per second (cps) exceeds 60 k counts. Compared with a flat surface, we conclude that the enhancement in photon collection efficiency is greater than 1.5.

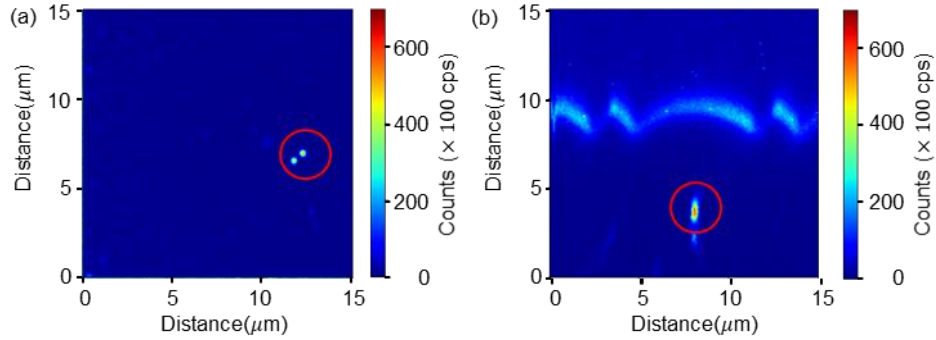

**Figure S3. (a)** and **(b)** show confocal microscopy images (CFM) in the cross-sectional and lateral sectional planes, respectively, after milling the Fresnel-type SIL with the first strategy (Fig. 2a, 2b, and 2d).

### S4. Gated photon counting

To choose the detection time window, we measured the time-resolved photoluminescence depending on the initial spin state $m_s=0$ and $m_s = -1$. Figures S4. a S4. c are obtained from the NV⁻ center with the Fresnel-type SIL and under the flat surface, respectively. Figures S4. b and S4. d Ratio of photoluminescence from different spin states. We chose the 50 ns time window during which result the maximum contrast in spin dependent photon count.

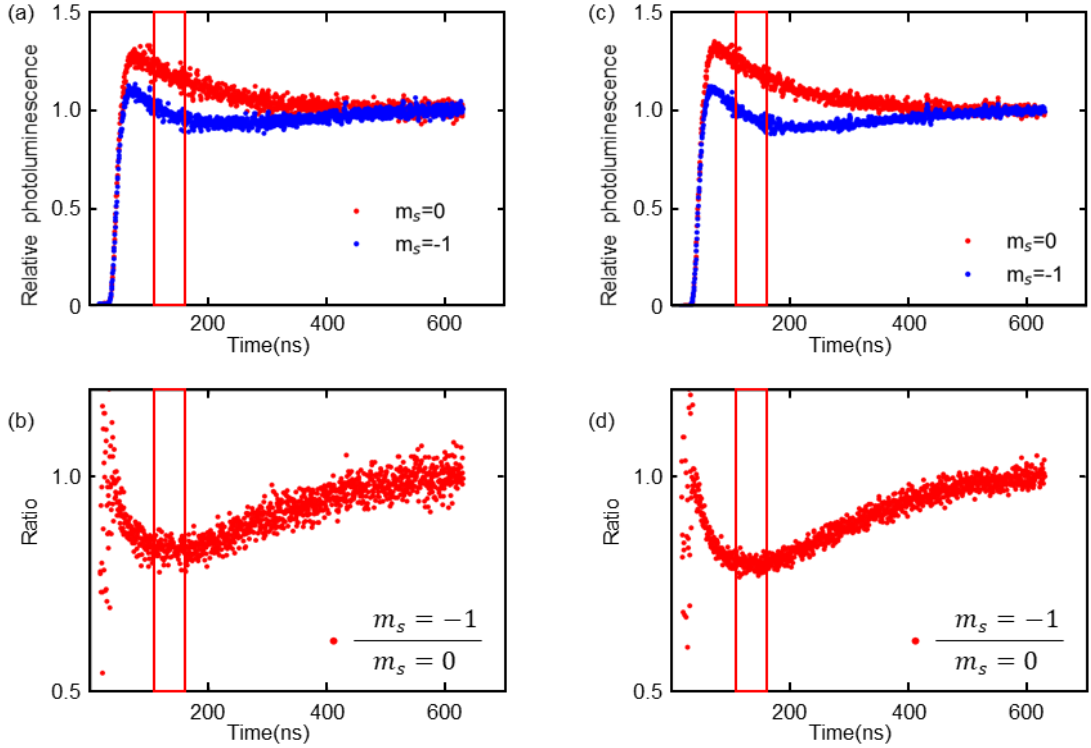

**Figure S4. (a)** The time-correlated single-photon detection data depending on the spin state ($m_s$=0 and $m_s = -1$) from the NV⁻ center with the Fresnel-type SIL, and **(b)** shows the ratio between the photon counts of the two spin states. **(c)** and **(d)** are the counterparts of **(a)** and **(b)** for the NV⁻ center under the flat surface. The red boxes represent a 50 ns width time windows.

### S5. Finite difference method simulation

To quantify the enhancement in photon collection efficiency, we simulated the finite difference method. First, we simplified this problem to a single-frequency problem (1) and used symmetry in the azimuthal angle (2). From (1), we can use the Helmholtz equation with two wavenumbers $k$ (wavenumbers in diamond and oil). From (2), we obtain $\frac{\partial V}{\partial \varphi} = 0$. The governing equation becomes:

$$\nabla^2 V + k^2 V = 0$$

$$\frac{1}{r^2}\frac{\partial}{\partial r}\left(r^2 \frac{\partial V}{\partial r}\right) + \frac{1}{r^2 \sin\theta}\frac{\partial}{\partial \theta}\left(\sin\theta \frac{\partial V}{\partial \theta}\right) + k^2 V = 0$$

$$\frac{\partial^2 V}{\partial r^2} + \frac{2}{r}\frac{\partial V}{\partial r} + \frac{1}{r^2}\frac{\partial^2 V}{\partial \theta^2} + \frac{1}{r^2 \tan\theta}\frac{\partial V}{\partial \theta} + k^2 V = 0$$

We input a *h* as finite difference in *the r* direction, and τ as a finite difference in *the θ* direction. The number of meshes was constrained to $0 \le i \le N_r$ and $0 \le j \le N_\theta$. Then we get

$$\left(-\frac{2}{h^2} - \frac{2}{r_i^2 \tau^2} + k_{ij}^2\right) V_{i,j} + \left(\frac{1}{h^2} - \frac{1}{r_i h}\right) V_{i-1,j} + \left(\frac{1}{h^2} + \frac{1}{r_i h}\right) V_{i+1,j}$$
$$+ \left(-\frac{1}{2\tau r_i^2 \tan\theta} + \frac{1}{r_i^2 \tau^2}\right) V_{i,j-1} + \left(\frac{1}{2\tau r_i^2 \tan\theta} + \frac{1}{r_i^2 \tau^2}\right) V_{i,j+1} = 0$$

And we simplify by changing the coefficients.

$$A_{i,j} V_{i,j} + B_{i,j} V_{i-1,j} + C_{i,j} V_{i+1,j} + D_{i,j} V_{i,j-1} + E_{i,j} V_{i,j+1} = 0$$

In the Sommerfeld radiation condition, the NV⁻ center is the emitting source; therefore, at the boundary, it is just outgoing.

$$\left(\frac{1}{c^2}\frac{\partial^2}{\partial t^2} - \nabla^2\right) U = 0$$

$$\left(\frac{1}{c}\frac{\partial}{\partial t} - \nabla\right)\left(\frac{1}{c}\frac{\partial}{\partial t} + \nabla\right) U = 0$$

$$\left(\frac{1}{c}\frac{\partial}{\partial t} + \nabla\right) U = 0$$

With $U(r,t) = V(r) e^{iwt}$,

$$\frac{\partial V}{\partial r} + ikV = 0$$

Again we input *h*,

$$V_{i+1,j} = V_{i-1,j} - 2ihkV_{i,j}$$

This equation is suitable at $i = N_r - 1$, so

$$A_{N_r-1,j}V_{N_r-1,j} + B_{N_r-1,j}V_{N_r-2,j} + C_{N_r-1,j}V_{N_r,j} + D_{N_r-1,j}V_{N_r-1,j-1} + E_{N_r-1,j}V_{N_r-1,j+1} = 0$$

$$V_{N_r,j} - V_{N_r-2,j} + 2ihkV_{N_r,-1j} = 0$$

From these two equations,

$$(A_{N_r-1,j} - 2ihkC_{N_r-1,j})V_{N_r-1,j} + (B_{N_r-1,j} + C_{N_r-1,j})V_{N_r-2,j} + D_{N_r-1,j}V_{N_r-1,j-1} + E_{N_r-1,j}V_{N_r-1,j+1} = 0$$

This was the boundary condition. By obtaining the eigenstate from these equations, we obtain a numerical solution. To solve the FDM, we assigned a material setting ($k$). In Fig. S5. a, the region colored yellow is diamond, and the other (dark blue) is oil. From this condition, we obtained the square of the field (Fig. S5. b), and is proportional to the photon number.

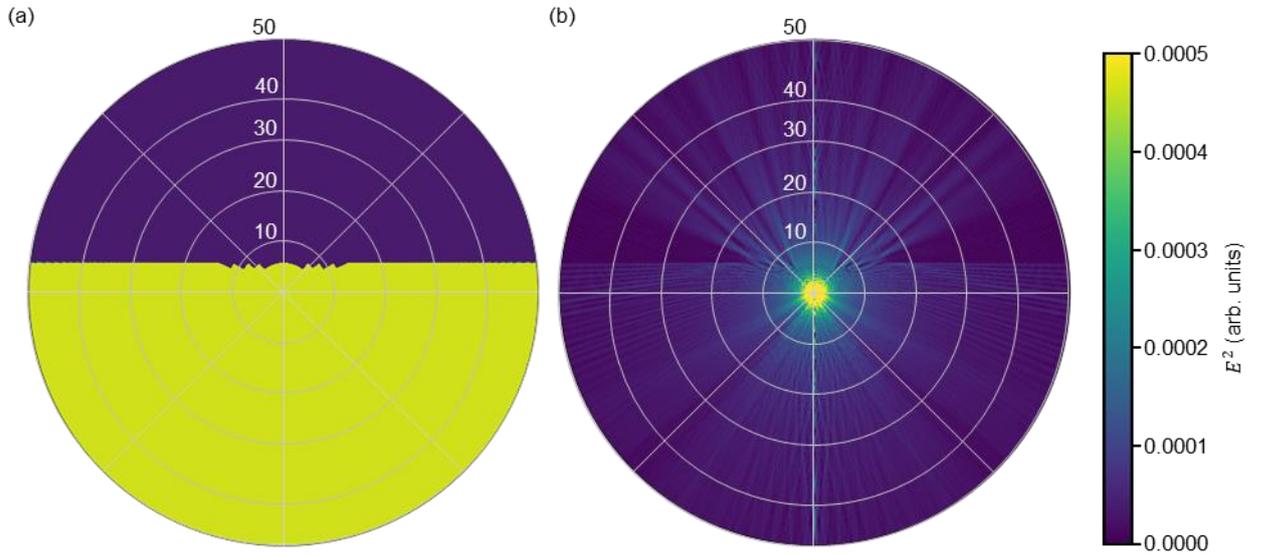

**Figure S5. (a)** shows the setup of the material. The yellow region represents the diamond material, and the dark blue region represents the oil for the objective lens. **(b)** shows value of Square of the electric field in an arbitrary color bar.

## Supporting Information reference